\documentclass[aps,prb,twocolumn,superscriptaddress,showpacs,showkeys,letterpaper]{revtex4}
\usepackage{etex} 
\usepackage{amsmath}
\usepackage{amssymb}
\usepackage{dsfont}
\usepackage[all,arc,knot,web]{xy}
\SelectTips{lu}{12}
\usepackage{graphicx}
\usepackage{hyperref}
\usepackage{color}
\usepackage{xcolor}
\definecolor{Blue}{rgb}{0.3,0.3,1}
\usepackage{pgffor}
\usepackage{verbatim}
\usepackage{tikz}
\usepackage{wrapfig}

\usetikzlibrary{arrows,matrix,decorations.markings}
\usetikzlibrary{positioning}
\usetikzlibrary{decorations.markings}
\usetikzlibrary{decorations.pathmorphing}
\tikzset{->-/.style={decoration={markings,mark=at position #1 with {\arrow{>}}},postaction={decorate}}}
\tikzset{-<-/.style={decoration={markings,mark=at position #1 with {\arrow{<}}},postaction={decorate}}}
\newcommand{\unit}{\mathbf{1}}
\newcommand{\Z}{\mathbb{Z}}
\newcommand{\J}{\mathcal{J}}
\newcommand{\ket}[1]{\left|{#1}\right\rangle}
\newcommand{\bra}[1]{\left\langle{#1}\right|}
\newcommand{\bmm}{\begin{matrix}}
\newcommand{\emm}{\end{matrix}}
\newcommand{\upperRomannumeral}[1]{\uppercase\expandafter{\romannumeral#1}}

  \newcommand{\TriangleYpart}[6]{
  	\def\jjA{#1}
  	\def\jjB{#2}
  	\def\jjC{#3}
  	\def\jjD{#4}
  	\def\jjE{#5}
  	\def\jjF{#6}
  	\TriangleYpartExtended
  }
   \newcommand{\TriangleYpartExtended}[6]{
   	\bmm\xy
   	0;/r0.12pc/:;
   	(-10,10)*{}="p1";
   	(10,10)*{}="p2";
   	(0,-10)*{}="p3";
   	(-4,5)*{}="p4";
   	(4,5)*{}="p5";
   	(0,-3)*{}="p6";
   	"p4";"p1" **\dir{-} ?(0.6)*\dir{#5}+(-3,-1) *{\scriptstyle \jjE};
   	"p5";"p2" **\dir{-} ?(0.6)*\dir{#6}+(3.5,-1) *{\scriptstyle \jjF};
   	"p3";"p6" **\dir{-} ?(0.6)*\dir{#1}+(3,-1) *{\scriptstyle \jjA};
   	"p4";"p5" **\dir{-} ?(0.5)*\dir{#4}+(0,3.5) *{\scriptstyle \jjD};
   	"p6";"p5" **\dir{-} ?(0.5)*\dir{#3}+(3.5,0) *{\scriptstyle \jjC};
   	"p6";"p4" **\dir{-} ?(0.5)*\dir{#2}+(-3,0) *{\scriptstyle \jjB};
   	\endxy\emm
   }

\makeindex

\begin{document}

\title{On Quantum Entanglement in Topological Phases on a Torus}
\author{Zhu-Xi Luo}
\author{Yu-Ting Hu}
\affiliation{Department of Physics and Astronomy, University of Utah, Salt Lake City, Utah, 84112, U.S.A.}
\author{Yong-Shi Wu}
\affiliation{Key State Laboratory of Surface Physics, Department of Physics and Center for Field Theory and Particle Physics, Fudan University, Shanghai 200433, China}
\affiliation{Collaborative Innovation Center of Advanced Microstructures, Nanjing 210093, China} 
\affiliation{Department of Physics and Astronomy, University of Utah, Salt Lake City, Utah, 84112, U.S.A.}

\date{\today}

\begin{abstract}
In this paper we study the effect of non-trivial spatial topology on quantum entanglement by examining the degenerate ground states of a topologically ordered system on torus. Using the string-net (fixed-point) wave-function, we propose a general formula of the reduced density matrix when the system is partitioned into two cylinders. The cylindrical topology of the subsystems makes a significant difference in regard to entanglement: a global quantum number for the many-body states comes into play, together with a decomposition matrix $M$ which describes how topological charges of the ground states decompose into boundary degrees of freedom. We obtain a general formula for entanglement entropy and generalize the concept of minimally entangled states to minimally entangled sectors. Concrete examples are demonstrated with data from both finite groups and modular tensor categories (i.e., Fibonacci, Ising, etc.), supported by numerical verification. 
\end{abstract}


\maketitle
\section{Introduction}\label{Intro}

Quantum phases of matter with intrinsic topological order\cite{wen1995topological,nayak2008non} are exotic gapped states that cannot be described in Landau's symmetry-breaking regime. They are characterized by novel properties, such as topological degeneracy of ground states\cite, fractional quantum numbers and fractional statistics of excitations and bulk-boundary correspondence. By now it is widely accepted that the existence of long-range entanglement in these phases is the physical origin of these exotic topological properties. Therefore entanglement measurements are powerful tools in understanding or characterizing topological phases. Topological entanglement entropy\cite{Preskill,WenTEE,Balents} and entanglement spectrum\cite{Spectrum} of the ground states are known to partially characterize intrinsic topological order, while minimally entangled ground states\cite{Ashvin} can be used to extract quasi-particle statistics of the system. Further, entanglement entropy can be used in detecting both symmetry protected and symmetry enriched topological phases\cite{Pollman,Huang,Lu,Charged} as well. 

To study how non-trivial spatial topology of a two-dimensional system in a topological phase affects quantum entanglement, we concentrate on the string-net ground state wave functions. String-net models\cite{StringNet} are exactly soluble lattice-spin models defined on a trivalent graph which realize a large class of non-chiral 2+1D topological phases. These include all phases whose low energy effective theories are discrete gauge theories or doubled Chern-Simons theories.  String-net states have universal significance when dealing with topological phases: They are some typical tensor network states\cite{ChenGuWen}, which can be more generally viewed as renormalization group fixed-point states or wave functions\cite{ChenGuWen2, Vidal} because of their topological invariance\cite{HuStirlingWu}. 

The set of input data to define the model is $\left\{I,d,\delta,G\right\}$. $I=\left\{0,1,\cdots,N-1\right\}$ specifies $N$ string types that are placed on each oriented link in the graph. For each $j\in I$, there is a dual string type $j^*$ corresponding to reversed orientation of the link. $d_j=d_{j^*}$ is the quantum dimension assigned to each string type $j\in I$. The rank-three tensor $\delta_{ijk}$ specifies fusion rules, i.e., the allowed branchings of the graph. (For simplicity we assume the multiplicity-free case and take $\delta_{ijk}$ to be either zero or one.) $G_{kln}^{ijm}$ are the symmetrized $6j$-symbols with $i,m,k,l,n\in I$ satisfying certain algebraic relations. This set of data defines a unitary fusion category $\mathcal{C}$. For a review of the Hamiltonian of the model, see Appendix \ref{SN}.

Given a fixed bipartition of the graph into two parts $A$ and $B$, one can calculate for a state $\Psi$ of the system the density matrix $\rho_{AB}=\ket{\Psi}\bra{\Psi}$ and the reduced density matrix $\rho_A=Tr_B(\rho_{AB})$. The entanglement entropy is then given by $S=-Tr\left(\rho_A\log \rho_A\right)$. For string-net ground states with $A$ a simply-connected region (disk, sphere, etc.), it is known\cite{WenTEE} to be
\begin{equation}
\label{eq:disk}
S=-L\sum_{j\in I} \frac{d_j^2}{D}\log \frac{d_j}{D}-\gamma,
\end{equation}
where $L$ is the number of links along the boundary of the cut,  $D=\sum_{j\in I} d_j^2$ the total quantum dimension and $\gamma=\log D$. 

The first term demonstrates the familiar area law, while the second term is independent of $L$ and vanishes when the system does not have topological order ($D=1$). This gives rise to topological entanglement entropy $\gamma$ which is non-local and invariant under smooth deformation of the ground state. 

On a nontrivial manifold such as torus, systems with intrinsic topological order can possess degenerate ground states. Furthermore, nontrivial topologies of the two subregions will affect the entanglement entropy.\cite{Yao,Fradkin,Ashvin} The total quantum dimension $D$ is then unable to distinguish between those degenerate ground states. This is exactly the point where entanglement entropy enters the game. In chiral Chern-Simons theories on torus, it is shown by the surgery method\cite{Witten,Fradkin} that topological entanglement entropy depends on linear combinations of basic degenerate ground states. 

For non-chiral string-net models on torus, there is no general analytic result considering two cylindrical subsystems $A,B$.

In this work, we propose an expression for general reduced density matrices in regard to the cylindrical bipartition and show that they are sensitive to different topological charges $\J$ of ground states $\psi_\J$ of the model. Furthermore, an additional decomposition matrix $M$ enters the expression, which describes how bulk topological charges of the ground states split into boundary degrees of freedom upon taking the partial trace. The specific form of reduced density matrix makes it natural to define the notion of entanglement sectors, which are sets of degenerate ground states that have the same decomposition pattern. Making use of the reduced density matrices, we derive a general formula for entanglement entropy. The $L$-dependent term in the expression for entanglement entropy have similar structures as that on a disk-shaped region, while the topological ($L$-independent) entanglement entropy is significantly different. Variation shows that the entanglement entropy is minimized when the state contains only linear superpositions of ground states in the same entanglement sector. Furthermore, we verify that the upper bound of topological entanglement entropy is consistent with the strong sub-additivity property on torus derived in ref.\cite{Ashvin}. We propose that above properties hold generally for topological phases. Additionally, we discuss the applications of our results to entanglement spectra and R\'enyi entropies.

The paper is organized as follows. In section \ref{Disk}, we review the derivation of eq.\eqref{eq:disk} and motivate our new formulation of reduced density matrices. In section \ref{Torus} we take care of the case where both subsystems are cylinders and give the general formula for entanglement entropy. Section \ref{Exam} focuses on concrete models constructed from irreducible representations of $\Z_n$ groups, (non-Abelian) $S_3$ group and modular tensor categories, i.e., the double semion, double Fibonacci, double Ising model, etc.. Section \ref{Summary} discusses the implications on entanglement spectra, R\'enyi entropies, boundary theories and outlooks future directions.

\section{Entanglement Entropy on a Disk-shaped Region}\label{Disk}

In this section, we reformulate the expression of reduced density matrix\cite{WenTEE} on a disk-shaped region, which will be convenient to generalize to regions with nontrivial topology. 

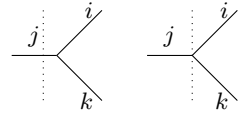
\begin{figure}[htbp]
\centering
\begin{tikzpicture}[scale=0.6]
\draw (3,0)--(4,0);
\draw (4,0)--(5,1);
\draw (4,0)--(5,-1);
\node[above] at (3.5,0) {$j$};
\node[above,left] at (5,1) {$i$};
\node[below,left] at (5,-1) {$k$};
\draw[dotted] (4,1)--(4,-1);
\draw (0,0)--(1,0);
\draw (1,0)--(2,1);
\draw (1,0)--(2,-1);
\node[above] at (0.5,0) {$j$};
\node[above,left] at (2,1) {$i$};
\node[below,left] at (2,-1) {$k$};
\draw[dotted] (0.7,1)--(0.7,-1);
\end{tikzpicture}
\caption{Left: rough cut. The cut (dotted line) intersects some link $j$. Right: smooth cut. There is no intersection.}
\label{fig:rough}
\end{figure}

Consider a bipartition of the trivalent graph on an \textit{arbitrary} manifold into two subsystems $A,B$ with $A$ having the shape of a disk. For simplicity, on a trivalent graph we always take the cuts which bipartites the system to be those intersecting some links in the middle, corresponding to the left figure in Fig.\ref{fig:rough}. 

\begin{figure}[htbp]
\centering
\begin{tikzpicture}[scale=0.8]
\draw[-<-=.5] (0,0)--(1,0);
\node[below] at (0.5,0) {$p_1$};
\draw[->-=.5] (1,0)--(2,0);
\node[below] at (1.5,0) {$s_1$};
\draw[->-=.5] (2,0)--(3,0);
\node[below] at (2.5,0) {$s_2$};
\draw (3,0)--(5,0);
\draw[->-=.5] (5,0)--(6,0);
\node[below] at (5.5,0) {$p_L$};
\draw[->-=.5] (0,0)--(0,1);
\draw[->-=.5] (1,0)--(1,1);
\draw[->-=.5] (2,0)--(2,1);
\draw[->-=.5] (3,0)--(3,1);
\draw[->-=.5] (5,0)--(5,1);
\draw[->-=.5] (6,0)--(6,1);
\node[above] at (0,1) {$p_1$};
\node[above] at (1,1) {$p_2$};
\node[above] at (2,1) {$p_3$};
\node[above] at (3,1) {$p_4$};
\node[above] at (5,1) {$p_{L-1}$};
\node[above] at (6,1) {$p_L$};
\node at (4,0.5) {$\cdots$};
\draw[dashed] (6,0) to[out=315,in=225] (0,0);
\node[above] at (3,-1.3) {$0$};
\end{tikzpicture}
\caption{Basic tree-like string-net configuration in a disk-shaped region. $p_1,\cdots,p_L$ are boundary links, while $s_1,\cdots,s_{L-3}$ are interior links.}
\label{fig:disk}
\end{figure}
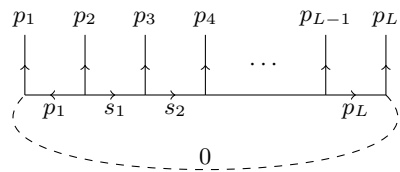

Following ref.[\onlinecite{WenTEE}], any string-net configuration inside a disk-shaped region can always be deformed\cite{StringNet} to basic tree-like configurations in fig.\ref{fig:disk}. Outside the disk-shaped region, any configurations can be deformed to similar tree-like configurations. 

Tracing out the degrees of freedom in $B$, the reduced density matrix for the subsystem $A$ is, (see eq. (9) of ref. [\onlinecite{WenTEE}]), 
\begin{equation}
\label{eq:WenRDM}
\langle \left\{ p', s'\right\}\mid \rho_L \mid \left\{p, s\right\}\rangle = \delta_{\{p\},\{p'\}}\delta_{\{s\},\{s'\}}\prod_{m=1}^L d_{p_m}
\end{equation}
up to a normalization factor of $D^{1-L}$, with $L$ the number of links at the boundary. 

One observes that in the above eq.\eqref{eq:WenRDM}, the reduced density matrix is diagonal and depends only on quantum dimensions of boundary links. 

Specifically, consider the case where we have three boundary links. For a possible $L=3$ configuration, the admissibility condition or branching rule $\delta_{p_1p_2p_3}=1\nonumber$ must be satisfied.
The reduced density matrix is thus
\begin{equation}
\rho_3=\frac{1}{D^2}\mathop{\oplus}\limits_{p_1,p_2,p_3} \left(\delta_{p_1p_2p_3}d_{p_1}d_{p_2}d_{p_3}\right), \nonumber
\end{equation}
with $x\oplus y$ understood as the $2\times 2$ diagonal matrix with diagonal entries $x,y$.

Similarly, in the $L=4$ case,  the branching rules $\delta_{p_1p_2s_1}=\delta_{s_1^*p_3p_4}=1\nonumber$
must be satisfied. The reduced density matrix is then
\begin{equation}
\rho_4=\frac{1}{D^3}\mathop{\oplus}\limits_{s_1,p_1,p_2,p_3,p_4} \left(\delta_{p_1p_2s_1}\delta_{s_1^*p_3p_4}d_{p_1}d_{p_2}d_{p_3}d_{p_4}\right).\nonumber
\end{equation}

One observe that both the number of $\delta$'s and $d$'s increase with respect to $L$. Since we are interested in the large $L$ limit, it would be convenient to reformulate the above expression in a more compact form. Furthermore, we want $L$ to be explicit in the equation. These considerations lead naturally to the formulation of reduced density matrix using fiber fusion category(FFC). We do not require most of the machinery of FFC for the formulation and will only list the relevant, easy-to-understand pieces. For readers who are mathematically intended, the concepts of FFC is briefly reviewed in Appendix \ref{app:Fiber}. 
 
Define $\alpha$ to be a complex function of variable $j\in I$ with values in complex numbers $\mathbf{C}$,
\begin{equation}
\alpha(j)=\alpha_j= d_j/D.
\end{equation}  
We define a tensor product by,
\begin{equation}
\alpha^{\otimes 2}_k\equiv \left(\alpha\otimes \alpha\right)_k=\mathop{\oplus}\limits_{i,j\in I}\left(\alpha_i\cdot \alpha_j\right)\delta_{ijk^*}.
\end{equation} 
This is a diagonal matrix, i.e., $x\oplus y$ is again understood as a $2\times 2$ matrix whose nonzero entries are the diagonal ones $x,y$.

One step forward, we have 
\begin{equation}
\alpha^{\otimes L}_k\equiv\left(\alpha^{\otimes (L-1)}\otimes \alpha\right)_k=\mathop{\oplus}\limits_{i,j\in I}\left(\alpha^{\otimes (L-1)}_i\cdot \alpha_j\right)\delta_{ijk^*}.
\end{equation}

For $L=3$ one has,
\begin{equation}
\alpha^{\otimes 3}_k=\mathop{\oplus}\limits_{i,j,l,m\in I}\left[(\alpha_i\alpha_j\delta_{ijl^*})\alpha_m \delta_{lmk^*}\right].\nonumber
\end{equation}

Plugging in the definition of $\alpha$ and relabel $i,j,m$ as $p_1,p_2,p_3$, this reduces to
\begin{equation}
\alpha^{\otimes 3}_k=\frac{1}{D^3}\mathop{\oplus}\limits_{p_1,p_2,l,p_3\in I}\left(\delta_{p_1p_2l^*}\delta_{lp_3k^*}d_{p_1}d_{p_2}d_{p_3}\right).\nonumber
\end{equation}
Comparing the above expression with that of $\rho_3$, we observe that we have one additional $\delta$ to get rid of. To this end, we introduce a projection operator $P_0$, which maps any vector $f=\left\{f_0, f_1,\cdots,f_{N-1}\right\}$ to $P_0(f)=\left\{f_0,0,\cdots,0\right\}$. (Components of the vector $f$ does not have to be complex numbers; they are generally matrices. Projecting out a matrix thus sets all its entries to be zero.) Graphically, this projection corresponds to the link $0$ in fig.\ref{fig:disk}. Applying $P_0$ to the above equation, we have
\begin{equation}
P_0 \left(\alpha^{\otimes 3}\right)=\frac{1}{D^3}\mathop{\oplus}\limits_{p_1,p_2,p_3} \left(\delta_{p_1p_2p_3}d_{p_1}d_{p_2}d_{p_3}\right),
\end{equation}
where the identities $d_j=d_{j^*}$ and $\delta_{lp_30}=\delta_{p_30l^*}$ for all $j\in I$ have been used.  (Here the $\delta$-symbol on the right hand side is understood as the Kronecker delta). We arrive at the equality $\rho_3 = 
DP_0\left(\alpha^{\otimes 3}\right)$. 

Generally we write the reduced density matrix
\begin{equation}\label{diskRDM}
\overline{\rho}_L=D P_0\left(\alpha^{\otimes L}\right).
\end{equation}
(The overline on $\rho$ is a notation instead of conjugation.) One can easily work out the case for $L=4$, and see that $\overline{\rho_4}$ coincide with the $\rho_4$ given above. Before stating a general proof, we note that $\alpha^{\otimes L}$ is fully determined by boundary degrees of freedom, while the projection enforces a global constraint on the system. We expect this independence of bulk degrees of freedom to be a general feature of disk-shaped systems with topological order, as partially implied in references [\onlinecite{Ludwig,Hamma1,Hamma2,WenRDM}].

\textit{Theorem \upperRomannumeral{1}}. Entanglement entropy calculated from the density matrix \eqref{diskRDM} is the same as from eq.\eqref{eq:disk}.

\textit{Lemma (1)}. Denote $\beta=\alpha^{\otimes L}$ with $L$ a positive integer and $\beta_j=P_j(\beta)$ with $P_j$ the projection operator mapping any vector of matrices $f=\left\{f_0,\cdots,f_j,\cdots,f_{N-1}\right\}$ to $P_j(f)=\left\{0,\cdots,f_j,\cdots,0\right\}$. Then
\begin{equation}\label{eq:Lemma1}
\frac{D}{d_j} tr \beta_j =1.
\end{equation}
Note that a sepcial case of the lemma with $j=0$ guarantees the trace of $\overline{\rho}_L$ to be unity. Proof can be found in Appendix \ref{app:Lemma}.

\textit{Lemma (2)}. \begin{equation}\label{eq:Lemma2}
tr\left(D\beta_0\log\left(D\beta_0\right)\right)=tr\left(\mathop{\oplus}\limits_{i} \frac{d_i}{D}\left(D\beta_i\right)\log\left(D\beta_i\right)\right)
\end{equation} 
Proof can be found in Appendix \ref{app:Lemma}.

\textit{Proof of Theorem \upperRomannumeral{1}}. Notice that to cut out a simply connected region, the number of links crossing the boundary between the two regions should satisfy $L\geq 2$. (One can also have the case $L=0$, which is of no physical interest and thus ignored.) 

For $L=2$, eq.\eqref{eq:disk} gives\begin{equation}
\begin{split}
S_2&=-\sum_k \frac{d_k^2}{D}\log\frac{d_k^2}{D}.
\end{split}\nonumber
\end{equation}

On the other hand, $\overline{S}_2 =-\overline{\rho}_2\log\overline{\rho}_2$ gives \begin{equation}
\begin{split}
\overline{S}_2&=-\left[DP_0\left(\alpha\otimes\alpha\right)\right]\log \left[DP_0\left(\alpha\otimes\alpha\right)\right]=S_2.
\end{split}\nonumber
\end{equation}

So we conclude for $L=2$, it is indeed the case that the entanglement entropies computed from these two methods are the same. For $L>2$, we just need to verify that \begin{equation}
\label{eq1}
\overline{S}_{L+1}-\overline{S}_L = -\sum_k\frac{d_k^2}{D}\log\frac{d_k}{D}.
\end{equation}

Consider the first term on the left hand side. Denoting $\beta=\alpha^{\otimes L}$ and using Lemma (1), one arrives at

\begin{equation}
\begin{split}
\overline{S}_{L+1}=-\sum_i\frac{d_i^2}{D}\log\frac{d_i}{D}-tr\left(\mathop{\oplus}\limits_{i}  \frac{d_i}{D}\left(D\beta_i\right)\log\left(D\beta_i\right)\right).\nonumber
\end{split}
\end{equation}
Thus \begin{equation}
\begin{split}
\overline{S}_{L+1}-\overline{S}_L=&-\sum_i\frac{d_i^2}{D}\log\frac{d_i}{D}+tr\left(D\beta_0\log\left(D\beta_0\right)\right)\\
&-tr\left(\mathop{\oplus}\limits_{i}   \frac{d_i}{D}\left(D\beta_i\right)\log\left(D\beta_i\right)\right).
\end{split}
\end{equation}
Then eq.\eqref{eq1} is verified using Lemma (2). $\square$

\section{Entanglement Entropy on a Torus}\label{Torus}
Topologically ordered systems can have degenerate ground states on a torus, i.e., energy itself is not able to distinguish among them. Entanglement entropy, on the other hand, serves as a quantitative measure to split the degeneracy from an information perspective, at least partially. In this section, we study nonchiral topological phases and give a general expression for both the reduced density matrix and entanglement entropy. It is argued in ref.\cite{Ashvin} that the ground states of a topologically ordered system with minimal entanglement entropy (with respect to the same bipartition) are eigenstates of Wilson loop operators defined for loops parallel to the cut. Below we show that for \textit{nonchiral} topological systems, the concept of minimally entangled states needs to be generalized to minimally entangled sectors (MESe): they are linear superpositions of specific minimally entangled states chosen in a unique way. Entanglement entropy cannot distinguish between states in the same sector.

\subsection{Ground State Wavefunction}
Consider any two trivalent graphs discretizing the same surface (torus here). Mathematically, they can always be mutated to each other by a composition of the local Pachner moves\cite{Pachner}. Physically, between the two (different) Hilbert spaces defined on the two graphs, there is a mutation transformation by a composition of the basic Pachner moves. It can be proved\cite{Thesis,StringNet} that the mutations are unitary in the ground-state subspace and the ground states are invariant under such mutations. An outline of the proof can be found in Section \upperRomannumeral{3} of ref.\cite{Full}. In other words, topological properties of ground states are robust against local deformations of the graph/Hilbert space.

As a result, we can always reduce an arbitrary graph on the torus to a simplest one using a sequence of Pachner moves and study the properties of ground states in a reduced Hilbert space. The other way around, we can always return to more complicated graphs easily using the mutation transformations. The simplest graph on a torus has two vertices, three edges and one plaquette, as indicated in the figure below. 

\begin{figure}[htbp]
\centering
\begin{tikzpicture}
\draw [blue, thick] (0,0) circle [x radius=2cm, y radius=1.2cm];
\draw (0,0) circle [x radius=1.5 cm, y radius=0.7 cm];
\draw [blue, thick] (1,0.1)  arc[x radius = 1 cm, y radius = 0.4 cm, start angle= 0, end angle= -180];
\draw [blue, thick] (0.7,-0.2) arc[ x radius = 0.7 cm, y radius = 0.4 cm, start angle=0, end angle=180];
\draw (-1.3,0.35) .. controls (-1.15,1) and (-1.1,1.1) .. (-1,1.05);
\draw [dashed] (-1,1.05) .. controls (-0.9,0.7) and (-0.4,0.2) .. (0,0.2);
\draw (0,0.2) .. controls (0.3,0.3) and (0.6,0.5) .. (0.7,0.62);
\node[left] at (-1.2,0.7) {$i$};
\node[left] at (0,0.9) {$k$};
\node[below] at (0,-0.7) {$j$};
\end{tikzpicture}
\caption{The simplest graph on torus contains two vertices, three links and one plaquette.}
\label{fig:sim}
\end{figure}
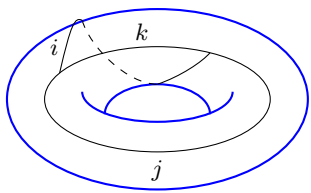

Any configuration on torus can thus be expressed on the simplest graph as $\ket{jik}$. Modular transformations on the ground-state subspace are constructed by,\cite{Thesis}\begin{equation}
\begin{split}
& \mathcal{S}\ket{jik}=\sum_{k'}v_k v_{k'}G_{jik'}^{j^*i^*k}\ket{ijk'},\\
& \mathcal{T}\ket{jik}=\sum_{k'}v_k v_{k'} G_{jik'}^{j^*i^*k'}\ket{jk'i},
\end{split}
\end{equation}
where we have chosen the $\mathcal{T}$-action to be cutting the uncontractible loop consisting of the links $j$ and $k$ (the meridional loop), twisting the boundary by $2\pi$ and gluing it back. We lay the ground states in the basis composed of the eigenvectors of $\mathcal{T}$,
\begin{equation}
\mathcal{T}\ket{\psi_\J}=\theta_{\mathcal{J}}\ket{\psi_\J}.
\end{equation}
This basis $\left\{\psi_\J\right\}$ is also called the quasiparticle basis in the literature. The symbol $\J$ labels species of topological charges (or equivalently bulk quasiparticles) and $\theta_\J$ is the twist of $\J$. 

The expression of $\ket{\psi_\J}$ satisfying the above equation is\cite{Thesis},
\begin{equation}\label{eq:gsJ}
\ket{\psi_\J}=\frac{1}{\sqrt{D}}\sum_{ijk}\frac{v_iv_k}{v_j}z^\J_{jijk}\ket{jik}.
\end{equation}
The $z^\J_{jijk}$ tensor is called the half-braiding tensor satisfying some consistency conditions that characterize each $\psi_\J$. Every such tensor is a map of vector spaces $V_k^{ji}\rightarrow V_k^{ij}$ which results from exchanging $i$ and $j$ which are in the $k$-fusion channel. An introduction to these half-braiding tensors can be found in ref.\cite{Full}, and we briefly review the notations in appendix \ref{app:HalfBraiding}. If the input category is the representation category of an Abelian group $\mathbb{Z}_N$ or a modular tensor category, the analytic expression for those $z$ tensors are known\cite{Full}. For non-Abelian groups a numerical solution is generally involved. 
 
Physically, every such state \eqref{eq:gsJ} can be understood as creating a pair of excitations with topological charges $\J$ and $\J^*$ from the trivial ground state, winding one of them around one uncontractible loop of the torus and annihilate them. In other words, a ground state $\psi_\J$ can be understood as a flux of quasiparticle $\J$ threading through the torus, see the fig.\ref{fig:cuts} below. 
 
The reason why those $z$-tensors enter the expression can be understood as follows. The most distinguished property of the excitations in a topological phase is that they can have nontrivial braiding statistics, arising from the half-braiding tensors $z$. These half-braiding tensors need to satisfy the naturality conditions\cite{Full} that guarantee the consistent properties of two sequential braidings. Each independent solution to the naturality conditions corresponds mathematically to an irreducible representation $\J$ of the quantum double of the input data. Therefore, each quantum double label $\J$ offers a consistent way for some potential excitation to braid and thus corresponds to one excitation species. Due to the correspondence between quasiparticles and ground states described in the last paragraph, it is natural that every ground state in quasiparticle basis can be explicitly constructed from the quantum double of the input data, i.e., the $z$-tensors.

We focus on the bipartition where both subsystems have cylindrical topologies. Our partition, named as double-cylinder cuts, is specified in fig.\ref{fig:cuts}. The two red cuts (color online) are parallel to the meridional loop of the torus, crossing $L_1$ and $L_2$ links respectively. Directions of the links are fixed as in the figure. When the configurations do not match the such directions, one can simply reverse orientation of those mismatched strings and relabel them as their dual strings.

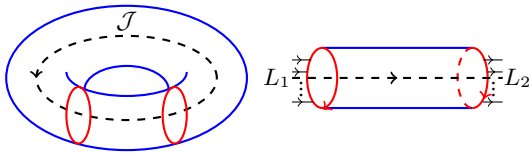
\begin{figure}[htbp]
\centering
\begin{tikzpicture}[scale=.8]
\draw [blue, thick] (0,0) circle [x radius=2cm, y radius=1.2cm];
\draw [thick, dashed,->-=0.5] (0,0) circle [x radius=1.5 cm, y radius=0.7 cm];
\node [above] at (0,0.65) {$\J$};
\draw [blue, thick] (1,0.1)  arc[x radius = 1 cm, y radius = 0.4 cm, start angle= 0, end angle= -180];
\draw [blue, thick] (0.7,-0.2) arc[ x radius = 0.7 cm, y radius = 0.4 cm, start angle=0, end angle=180];
\draw [red, thick] (-0.8,-0.62) circle [x radius=0.2cm, y radius=0.47cm];
\draw [red, thick] (0.8,-0.62) circle [x radius=0.2cm, y radius=0.47cm];
\draw [blue, thick] (3.25,0.5)--(5.75,0.5);
\draw [blue, thick] (3.25,-0.5)--(5.75,-0.5);
\draw [red, thick, -<-=0.3] (5.75,-0.5) arc (-90:90:0.25cm and 0.5cm);
\draw [red, thick, dashed] (5.75,0.5) arc (90:270:0.25cm and 0.5cm);
\draw [red, thick, -<-=0.8] (3.25,0) circle [x radius=0.25cm, y radius=0.5cm];
\draw [thick, dashed, ->-=0.5] (2.75,0)--(6.25,0);
\draw [->-=0.5] (2.75,0.1)--(3.0,0.1);
\draw [->-=0.5] (2.75,0.3)--(3.07,0.3);
\draw [->-=0.5] (2.75,-0.4)--(3.1,-0.4);
\node at (2.9,-0.3) {$\cdot$};
\node at (2.9,-0.1) {$\cdot$};
\node at (2.9,-0.2) {$\cdot$};
\draw [->-=0.5] (6,0.1)--(6.25,0.1);
\draw [->-=0.5] (5.93,0.3)--(6.25,0.3);
\draw [->-=0.5] (5.92,-0.4)--(6.25,-0.4);
\node at (6.1,-0.3) {$\cdot$};
\node at (6.1,-0.1) {$\cdot$};
\node at (6.1,-0.2) {$\cdot$};
\node at (2.5,0) {$L_1$};
\node at (6.5,0) {$L_2$};
\end{tikzpicture}
\caption{Left: Ground state $\psi_\J$ and cylindrical bipartition. The two red (color online) cuts are parallel to the meridional loop of the torus. Right: Two boundaries of the cylinder possess $L_1$ and $L_2$ boundary links respectively. Directions of the links are fixed as in the figure (when the configuration doesn't match the such directions, reverse orientation of those mismatched strings and relabel them as their dual strings). These directions and can alternatively be understood as  ``direction'' of the cut, as indicated by the red arrows.}
\label{fig:cuts}
\end{figure}

\begin{figure}[htbp]
\centering
\begin{tikzpicture}
\draw [Blue, thick] (0,0) circle [x radius=2cm, y radius=1.2cm];
\draw [color=black,-<-=.25] (0,0) circle [x radius=1.5 cm, y radius=0.7 cm];
\draw [color=black,-<-=.1] (0,0) circle [x radius=1.5 cm, y radius=0.7 cm];
\draw [color=black,-<-=.5] (0,0) circle [x radius=1.5 cm, y radius=0.7 cm];
\draw [color=black,-<-=.8] (0,0) circle [x radius=1.5 cm, y radius=0.7 cm];
\draw [Blue, thick] (1,0.1)  arc[x radius = 1 cm, y radius = 0.4 cm, start angle= 0, end angle= -180];
\draw [Blue, thick] (0.7,-0.2) arc[ x radius = 0.7 cm, y radius = 0.4 cm, start angle=0, end angle=180];
\draw [color=black,thick,-<-=.5] (-1.3,0.35) .. controls (-1.15,1) and (-1.1,1.1) .. (-1,1.05);
\draw [color=black,thick,dashed] (-1,1.05) .. controls (-0.9,0.7) and (-0.4,0.2) .. (0,0.2);
\draw [color=black,thick] (0,0.2) .. controls (0.3,0.3) and (0.6,0.5) .. (0.7,0.62);
\node[left] at (-1,0) {$j'$};
\node[left] at (-1.2,0.7) {$i$};
\node[left] at (0,0.9) {$k$};
\node[right] at (1,0.1) {$j'$};
\node[below] at (0,-0.7) {$j$};
\draw [thick, color=red] (-1.5,-1.1)--(-0.5,-0.1);
\draw [thick, color=red] (0.5,-0.1)--(1.5,-1.1);
\node[below] at (0.4,-1.2) {$\textcolor{red}{A}$};
\node[above] at (1.8,0.7) {$\textcolor{red}{B}$};
\end{tikzpicture}
\caption{Partition of the simplest graph on a torus into two subsystems $A,B$. Cuts are denoted by red lines.}
\label{fig:SimCut}
\end{figure}
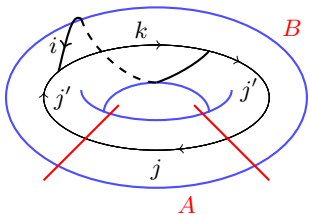

For the simplest graph described above, we have $L_1=L_2=1$. Subsystem $B$ consists of links $i,k$ and part of $j$, while subsystem $A$ contains the other part of $j$, as shown in figure \ref{fig:SimCut}.

Every configuration $\ket{jik}$ can then be written as,
\begin{equation}
\ket{jik}=\delta_{jj'}\ket{j'}_A\otimes\ket{jik}_B.
\end{equation}
(Note that because of this constraint expressed by Kronecker delta, smooth cuts which intersect with vertices instead of links and rough cuts which intersect with link are equivalent after taking the partial trace.)

Tracing out degrees of freedom in subsystem $B$, one obtain the reduced density matrix for the simplest graph as 
\begin{equation}
\begin{split}
\rho_A & = \frac{1}{D} \sum_j \ket{j}\sum_{i,k} \frac{d_id_k}{d_j} z^\J_{jijk} \overline{z^\J_{jijk}}\bra{j}  \\
& = \sum_j \ket{j}\frac{d_j}{d_\J} M_{\J j}\bra{j} ,
\end{split}
\end{equation}
where $d_\J$ is the quantum dimension for bulk quasiparticle $\J$ and we have defined a $M$ matrix as,
\begin{equation}
\label{eq:W}
\sum_{i,k} z^\J_{jijk} \overline{z^\J_{jijk}} \frac{d_id_k}{D d_j} = \frac{d_j}{d_\J} M_{\J j}.
\end{equation}
The matrix $M_{\J j}$ has its first subscript denoting different ground states (or equivalently different quasi-particles) $\J$. The second subscript denotes elements in the label set $I$. All the entries are non-negative integers. 

Physical meaning of the matrix will be illustrated in the next subsection.

\subsection{Reduced Density Matrix on a General Graph}
One can use the deformation laws\cite{Full} to obtain an arbitrary trivalent graph from the simplest one in fig.\ref{fig:sim}. Using the cuts defined above in fig.\ref{fig:cuts} on a general trivalent graph with the number of boundary links $L_1$ and $L_2$, one can trace out the degrees of freedom in one of the cylinders $B$ and calculate the diagonalized reduced density matrix of subsystem $A$ for $\psi_\J$ as,
\begin{equation}
\label{rdmJ}
\overline{\rho}_{A;\J}=\sum_{j\in I}\frac{d_j}{d_\J}M_{\J j} \left[ \frac{D}{d_j} P_j\left(\alpha^{\otimes L_1}\right) \times \frac{D}{d_j} P_j\left(\alpha^{\otimes L_2}\right)\right].
\end{equation}

The two factors in the square bracket in eq.\eqref{rdmJ} describe the two boundaries of the cylinder $A$, relatively independent as indicated by the direct product but subjected to one global constraint of conservation of quantum numbers given by $P_j$. The projection operators leave some of the entries of matrices $\alpha^{L_i}$ ($i=1,2$) to be zero, but do not eliminate the relevant rows or columns. Thus the summation here is the usual summation among matrices of same dimensions. $D/d_j$ is the normalization factor resulting from Lemma $2$. 

Introducing the $M$ matrix is one of our central ideas. It reflects how the information of different ground states are displayed on boundary degrees of freedom. Topological charges of ground states in topological phases are dyons, i.e., composites of charges and fluxes. One species of topological charge $\J$ can have one flux type and multiple charge types. To calculate the entanglement entropy, one needs to partition the system into $A$ and $B$, which creates by hand two trivial domain walls or boundaries between the two subsystems. After tracing out one of the subsystems $B$, information carried by $B$ ``collapses'' to the boundaries of $A$. 

Consider a pair of topological charges $\J$ and $\J^*$ created in the bulk of $A$ and moved to the two boundaries. Upon reaching the boundaries, they suffer from the partial trace; information of the fluxes are traced out while information of charges survives. Furthermore, charges are exactly described by objects in the label set $I$, i.e., they one-to-one corresponds to string types. Thus starting from a topological charge $\J$ in the bulk, one obtains some charges $j_1, j_2, \dots \in I$ expressed by the string types of boundary links. The fact that it is possible to have multiple $j$'s for a single $\J$ corresponds to the fact that a dyon species can carry multiple charges. 

\begin{figure}[htbp]
\centering
\begin{tikzpicture}
\draw [->-=.5] (2,0) arc (0:360: 2cm and 1cm);
\draw [-<-=.5] (1,-0.87)--(1,-1.87);
\node[right] at (1,-1.87) {$q_3$};
\node[left] at (1,-1.1) {$0$};
\draw [-<-=.5](1.5,-0.67)--(1.5,-1.67);
\node[right] at (1.5,-1.67) {$q_2$};
\node[right] at (1.45,-0.75) {$r_1$};
\draw [->-=.5](0.5,-0.95)--(0.5,0.05);
\node[right] at (0.5,0.05) {$p_4$};
\draw [->-=.5](0,-1)--(0,0);
\node[right] at (0,0) {$p_3$};
\node[left] at (0,-1.15) {$s_2$};
\draw [->-=.5](-0.5,-0.95)--(-0.5,0.05);
\node[right] at (-0.5,0.05) {$p_2$};
\node[left] at (-.5,-1.15) {$s_1$};
\draw [->-=.5](-1,-0.85)--(-1,0.15);
\node[right] at (-1,0.15) {$p_1$};
\node[left] at (-2,0) {$j$};
\draw [->-=.5] (1.9,-1.3)--(1.9,-0.3);
\node[right] at (1.9,-1.3) {$q_1$};
\end{tikzpicture}
\caption{Basic string-net configuration on a cylinder for $L_1=4$ and $L_2=3$. All the vertical boundary links point downward and all the links consisting the loop point in a counterclockwise direction. Links on the loop are labeled by $\left\{s\right\}$ and  $\left\{r\right\}$, while upper and lower tails are labeled by $\left\{p\right\}$ and  $\left\{q\right\}$.}
\label{fig:cylinder}
\end{figure}
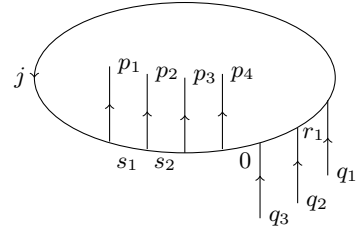

To be more concrete, we have
\begin{equation}
\J\rightarrow \mathop{\oplus}\limits_j M_{\J j} j.
\end{equation}
In other words, the entry $M_{\J j}$ is the multiplicity of the appearance of $j$ in the decomposition of $\J$. In the above fig.\ref{fig:cylinder}, one observes that the $M$ matrix simply describes how a ground state $\J$ is reflected on the link $j$ on the loop.

From the mathematical perspective, $M_{\J j}=\dim Hom(\J,j)$, which is the dimension of the homset of morphisms inside the input category $\mathcal{C}$. $\J$ is really an object in the category $\mathcal{Z}(\mathcal{C})$ which is the categorical (Drinfeld) center of $\mathcal{C}$. But we apply a forgetful functor to make $\J$ forget the half-braiding structure $z$ and $\J$ reduces to objects in $\mathcal{C}$. Thus calculation of entanglement entropy will be the decomposition map $\mathcal{F}:\mathcal{Z}(\mathcal{C})\rightarrow\mathcal{C}$. Since the input category $\mathcal{C}$ is the category of charges, what is left is the information of charges and the information of fluxes are lost.

We note that the above expression \eqref{rdmJ} for reduced density matrix can also be observed from procedures similar to that of part \ref{Disk}. For a cylinder with two boundaries, one can always reduce the string-net graph to one loop and $L_1+L_2$ tails, as shown in figure \ref{fig:cylinder}. The loop can be open if one erases the vacuum string $0$. Below is an example of $L_1=4, L_2=3$. 
 
\subsection{Entanglement Entropy}

For a general ground state $\Psi=\sum_{\J} c_\J \psi_\J$
with $c_\J$ complex numbers satisfying $\sum_{\J}\left|c_\J\right|^2=1$, the reduced density matrix is,

\begin{widetext}
\begin{equation}
\label{rdmMix}
\overline{\rho}_A= \sum_{\J} \left|c_\J\right|^2 \left\{\sum_{j\in I} \frac{d_j}{d_\J}M_{\J j} \left[ \frac{D}{d_j} P_j\left(\alpha^{\otimes L_1}\right)\times \frac{D}{d_j} P_j\left(\alpha^{\otimes L_2}\right) \right]\right\}.
\end{equation}
\end{widetext}

This expression is our main result. One can observe from the above expression that if two rows of the $M$ matrix are the same, i.e., if two $\J, \J'$ have the same pattern of decomposition into charges, then the reduced density matrix cannot distinguish between the them. We say that such ground states $\psi_\J$ and $\psi_{\J'}$ are in the same entanglement \textit{sector}. The reduced density matrix will be sensitive only to the summation of probabilities for states in the same entanglement sector, instead of the probability for every single $\psi_\J$.

It follows from eq.\eqref{rdmMix} and $S=-tr\left(\overline{\rho}_A \log \overline{\rho}_A\right)$ that the entanglement entropy is,
\begin{itemize}
\item[(1)] for $L_1=L_2=1$, the smallest possible graph for a cylinder, $S_{\left\{1,1\right\}} =\tilde{S}$, with  
\begin{equation}\label{S11}
\small{\tilde{S}=- \sum_j \left( \sum_{\J}\left|c_\J\right|^2 M_{\J j} \frac{d_j}{d_\J}\right) \log \left(\sum_{\J} \left|c_\J\right|^2 M_{\J j} \frac{d_j}{d_\J}\right)};
\end{equation}

\item[(2)] for $L_1=1,\ L=L_2>1$ or $L=L_1>1, \ L_2=1$, denoting 
\begin{equation}
\begin{split}
& a=-\sum_k \frac{d_k^2}{D} \log\frac{d_k}{D},\\
&b=\log D - \sum_{\J}\left|c_\J\right|^2 \log d_\J,\ \ \text{and}\\
& \tilde{S}_\J = - \sum_j \left(M_{\J j} \frac{d_j}{d_\J}\right)\log \frac{d_j}{d_\J},
\end{split}
\end{equation} the entanglement entropy is then \begin{equation}\label{S1L}
S_L=\ aL-b-\left(\sum_{\J} \left|c_\J\right|^2 \tilde{S}_\J - \tilde{S}\right).
\end{equation}

\item[(3)] for $L_1>1, L_2>1$, \begin{equation}\label{SL1L2}
\small{S_{\left\{L_1, L_2\right\}} =a(L_1+L_2-2)-b-\left(\sum_{\J} \left|c_\J\right|^2 \tilde{S}_\J -\tilde{S}'\right),}
\end{equation}
with $
\tiny{\tilde{S}'\equiv}  \small{-\sum_{i,j,k} \left[ \sum_\J \left| c_\J\right|^2 M_{\J j} \frac{d_j}{d_\J} \left(\frac{d_id_k}{Dd_j}\delta_{ikj^*}\right)\right]}\\ \times\small{\log \left[ \sum_\J \left| c_\J\right|^2 M_{\J j} \frac{d_j}{d_\J} \left(\frac{d_id_k}{Dd_j}\right)\right]}.$
\end{itemize}

The entanglement entropy for a single $\psi_\J$ in Case $(2)$ is \begin{equation}
\label{SJ}
S_{\left\{\J;L\right\}}=aL-\log D + \log d_\J.
\end{equation}
One observe that the $L$-dependent term contains $-\log D$, similar to the disk case\eqref{eq:disk}. This is due to the fact that, when $L_1$ or $L_2$ is equal to one, figure \ref{fig:cylinder} reduces to figure \ref{fig:disk} for the disk.  Furthermore, quantum dimension of the state $\psi_\J$ enters the above expression \eqref{SJ} as well.

Eq.\eqref{SL1L2} is our main interest and captures ``nonchiralness'' to the full. Observe that first term is now $a(L_1+L_2-2)$, indicating the reference point is now $(L_1, L_2)=(1,1)$ instead of $(0,0)$ since it is the smallest possible number of links that can contain the topological information of a torus. The second term is sensitive to specific ground state $\psi_\J$, while the third term is a measurement of the mixture of different ground states. 

Focusing on the $L$-independent part of eq.\eqref{SL1L2}, we write the topological entanglement entropy $\gamma$ as 
\begin{widetext}
\begin{equation}
S=aL-\gamma,\ \ \ \gamma=-\sum_k \frac{d_k^2}{D}\log\frac{d_k^2}{D}+2\log D-\sum_\J \left|c_\J\right|^2\log d_\J + \sum_\J \left|c_\J\right|^2 \tilde{S_\J} - \tilde{S}',
\end{equation}
\end{widetext}
which is maximized when the entanglement entropy is minimized. Comparing the above eq.\eqref{SL1L2} with \eqref{eq:disk}, one observes that the nontrivial topologies of the two subregions make a significant difference: apart from a global quantum number $\J$ which labels degenerate ground states, the decomposition matrix also comes into play.

Variations of $\gamma$ with respect to $\left|c_\J\right|^2$'s show that the point where all the first-order differentiations goes to zero realizes the minimal instead of maximal topological entanglement entropy. Maximum of the topological entanglement entropy can only be realized at points where the total probability of every entanglement sector is $0$ or $1$. So minimally entangled states generalize to minimally entangled sectors in our nonchiral case.

Depending on the input data, there exist possibilities where multiple entanglement sectors can maximize $\gamma$. But the entanglement sector which includes $\psi_\unit$, (where $\unit$ is the trivial quasiparticle or trivial ground state,) is definitely one of them. In other words, the case of $\J=\unit$ gives us the upper bound for topological entanglement entropy, which by straightforward calculation is $2\log D$. This is twice the topological entanglement entropy for the disk-shaped geometry, matching perfectly with the strong sub-additivity property (or called the uncertainty principle) of topological entanglement entropy on torus derived in ref.\cite{Ashvin}.

String-net models are exactly solvable models whose wavefunctions are fixed points of the renormalization group. More generally when one perturbs the wavefunctions away from string-net, we expect that there will be changes in the $L$-dependent term, but not in the $L$-independent term. The reason is that the $L$-independent term is topological and depends thus solely on the topological order of the system. Thus it will be invariant as long as the smooth deformation do not close the gap of the system.  

Further simplifications of the above equations will be in place when one considers Abelian models: the three different cases \eqref{S11},\eqref{S1L}, and \eqref{SL1L2} recombine to one expression.

\section{Examples}\label{Exam}
In this section, we study specific examples and compare entanglement entropy obtained from direct calculations and from the above formula. At least up to $L_1+L_2=9$, for all those examples the formula are exact. We argue that the formula is exact for all $L_1, L_2$. 

\subsection{$\Z_N$ Toric Code}
For $Rep_{\Z_N}$ model, the string types correspond to the $N$ irreducible representations of the group, denoted by $I=\left\{0,1,\dots,N-1\right\}$. Fusion rules are $\delta_{ijk}=1$ if $i+j+k=0 \mod N$ and $0$ if else. The $6j$ symbols are given by $
G^{ijm}_{kln}=\delta_{ijm}\delta_{klm^*}\delta_{jkn^*}\delta_{inl}.$

We have $N^2$ elementary ground states labeled by $\J=(g,\mu)$, with $g,\mu$ taking integer values $0,1,\dots,N-1$. $g$'s correspond to the group elements (fluxes), while $\mu$'s correspond to the irreducible representations (charges). The relevant decomposition matrix is given by \begin{equation}
M_{(g,\mu),j}=\delta_{\mu j}.
\end{equation}

For a simplest graph on torus, the ground states can be written as,
\begin{equation}
\ket{(g,\mu)}=\frac{1}{\sqrt{N}}\sum_{l,k\in I}e^{2\pi \mathrm{i} lg/N}\delta_{\mu lk^*} \ket{\mu lk}.
\end{equation}
 
For arbitrary $L_1, L_2>0$, the entanglement entropy is thus \begin{equation}\label{eq:Abelian}
\begin{split}
S=&(L_1+L_2-2)\log D \\
& -\sum_j \left(\sum_{\J} \left|c_\J\right|^2 M_{\J j}\right) \log \left(\sum_{\J} \left|c_\J\right|^2 M_{\J j}\right).
\end{split}
\end{equation}

Specifically, for the $\Z_2$ toric code model\cite{Toric}, the label set is $I=\left\{0,1\right\}$. All the strings are self-dual, i.e. $j=j^*$. The quantum dimensions are $d_0=d_1=1$ and the nontrivial fusion rule is $\delta_{110}=1$. Topological charges of ground states are $\J\in \left\{\unit,e,m,\epsilon\right\}$, the second of which often referred to as electric charge, the third the magnetic flux, while $\epsilon$ is a fermionic combination of the former two.

For the cuts defined above, the decomposition matrix is given by,
\begin{equation}
\label{M:Z2}
\begin{split}
& M_{\unit 0}=M_{m0}=1,\ M_{e0}=M_{\epsilon 0}=0,\\
& M_{\unit 1}=M_{m1}=0,\ M_{e1}=M_{\epsilon 1}=1.\nonumber
\end{split}
\end{equation}

The entanglement entropy is, for arbitrary $L_1>0, L_2>0$, \begin{equation}
\begin{split}
S= &(L_1+L_2-2)\log 2 \\
& -\sum_j \left(\sum_{\J} \left|c_\J\right|^2 M_{\J j}\right) \log \left(\sum_{\J} \left|c_\J\right|^2 M_{\J j}\right).
\end{split}
\end{equation}
If the ground state contains only a single $\psi_\J$, the above expression reduces to $S=(L_1+L_2-2)\log 2$.

\subsection{Double Fibonacci Model and General Modular Tensor Category Case}
Non-Abelianness introduces more interesting aspects into the system. We start with the simplest double Fibonacci model. The string types are $L=\{0,1\}$ 
(sometimes also denoted by $\{\mathbf{1},\tau\}$), both of which self-dual. Let $\phi=(1+\sqrt{5})/2$ be the golden ratio, the quantum dimensions are then $d_0=1$ and $d_1=\phi$.

The nonzero fusion rules are $\delta_{000}=\delta_{011}=\delta_{111}=1$. Nonzero independent $6j$-symbols $G$ are given by $G^{000}_{000}=1,\
G^{011}_{011}=G^{011}_{111}=1/\phi,\
G^{000}_{111}=1/\sqrt{\phi},\
G^{111}_{111}=-1/{{\phi}^2}.$ All other $6j$-symbols can be obtained from the Tetrahedral symmetry reviewed in Appendix \ref{SN}.

There are four quantum double labels: $\left\{0\overline{0},0\overline{1},1\overline{0},1\overline{1}\right\}$, or equivalently, $\left\{1\overline{1},1\overline{\tau},\tau\overline{1},\tau\overline{\tau}\right\}$ . The quantum dimensions are respectively $1,\phi,\phi,\phi^2$.

The decomposition matrix is,
\begin{equation}
\begin{split}
& M_{0\overline{0},0}=1,\ M_{0\overline{1},1}=1,\
M_{1\overline{0},1}=1,\\
& M_{1\overline{1},0}=1,\
M_{1\overline{1},1}=1.
\end{split}
\end{equation}

Ground states on the simplest graph are,
\begin{equation}
\begin{split}
\ket{0\overline{0}}&=\frac{1}{1+\phi^2}\left(\ket{000}+\phi\ket{011}\right);\\
\ket{0\overline{1}}&=\frac{1}{1+\phi^2}\left(\ket{101}+e^{4\pi i/5}\ket{110}-\sqrt{\phi}e^{2\pi i/5}\ket{111}\right);\\
\ket{1\overline{0}}&=\frac{1}{1+\phi^2}\left(\ket{101}-e^{\pi i/5}\ket{110}+\sqrt{\phi}e^{3\pi i/5}\ket{111}\right);\\
\ket{1\overline{1}}&=\frac{1}{1+\phi^2}\left(\ket{000}-\phi^{-1}\ket{011}+\ket{101}+\ket{110}+\phi^{-3/2}\ket{111}\right).
\end{split}
\end{equation}

We give a brief derivation of the entanglement entropy from the expression of reduced density matrix. Let's start from $\psi_{0\overline{0}}$. From eq.\eqref{rdmJ}, we have 
\begin{equation}
\rho_{\left\{A;0\overline{0};L_1,L_2\right\}}=D P_0 \left(\alpha^{\otimes L_1}\right)\times D P_0\left(\alpha^{\otimes L_2}\right).
\end{equation}

Denote $\alpha^{\otimes L_1}=\beta$, $\alpha^{\otimes L_2}=\sigma$. Using Lemma $(1)$ and Lemma $(2)$, it is easy to show that for $L_1\geq 2$,
\begin{equation}
S_{\left\{0\overline{0};L_1+1,L_2\right\}}-S_{\left\{0\overline{0};L_1,L_2\right\}}=a.
\end{equation}
Combining with the fact that the two boundaries of the cylinders are symmetric and that the cases where $L_1$ and $L_2$ are smaller than $2$ can be computed directly, one can prove the expression for entanglement entropy \eqref{S11}\eqref{S1L}\eqref{SL1L2}.

Relations for other $\J$'s can be proved in similar procedure. For a general ground state $\Psi=\sum_\J c_\J \psi_\J$, again we focus on the verification of 
\begin{equation}
S_{\left\{L_1+1,L_2\right\}}-S_{\left\{L_1,L_2\right\}}=a.
\end{equation}

Upon simplification, the left hand side becomes
\begin{equation}
\begin{split}
& \ a\left[\left(\left|c_{0\overline{0}}\right|^2+\phi^{-2}\left|c_{1\overline{1}}\right|^2\right)+\left(\left|c_{1\overline{0}}\right|^2+\left|c_{0\overline{1}}\right|^2+\phi^{-1}\left|c_{1\overline{1}}\right|^2\right)\right]\\
=&\ a=r.h.s..
\end{split}
\end{equation}

For general models constructed from modular tensor categories, including double semion and double Ising model, there is a general result. Consider the label set $I=\left\{0,1,\cdots,N-1\right\}$. We have for quantum doubles $N^2$ irreducible representations labeled by $\J=a\overline{b}$ with $a,b\in I$. The quantum dimensions are $d_{a\overline{b}}=d_ad_b$. The pure fluxes are those described by $b\overline{b}$. Decomposition matrices are given by\cite{Kong},
\begin{equation}
M_{a\overline{b},j}=\delta_{abj^*}.
\end{equation}

As additional examples, we carried out the direct computations of entanglement entropy for double semion and double Ising case up to $L_1+L_2=9$ and by comparison found the above formula for entanglement entropy exact.

\subsection{Non-Abelian Finite Group $S_3$}
Another interesting example concerns string-net model constructed from irreducible representations of non-Abelian finite groups $G$. The fluxes are classified using the conjugacy classes of the group, while the charges are classified by irreducible representations of the group. The elementary ground states and representations of quantum doubles are denoted by pairs $(A,\mu)$, where $A$ labels a conjugacy class of $G$, and $\mu$ is an irreducible representation of the centralizer $Z_A=\left\{g\in G\mid gh_A=h_Ag\right\}$. Here $h_A$ is an arbitrary representative in $C^A$ but is fixed once and for all. Note that $(A,\mu)$ reduces to $(g,j)$ in the Abelian case.

The simplest case is the $Rep_{S_3}$ model. The with label set $I=\left\{0,1,2\right\}$, quantum dimensions $d_0=d_1=1$ and $d_2=2$. Nontrivial fusion rules are $\delta_{110}=\delta_{220}=\delta_{221}=\delta_{222}=1$.

The independent nonzero symmetrized $6j$ symbols are,

\begin{align}
\label{S36js}
G^{000}_{000}=1,
G^{000}_{111}=1,
G^{000}_{222}=\frac{1}{\sqrt{2}},
G^{011}_{011}=1,
G^{011}_{222}=\frac{1}{\sqrt{2}},
\nonumber\\
G^{022}_{022}=\frac{1}{2},
G^{022}_{122}=\frac{1}{2},
G^{022}_{222}=\frac{1}{2},
G^{122}_{122}=\frac{1}{2},
G^{122}_{222}=-\frac{1}{2}.
\end{align}

All group elements of $S_3$ (or $D_3$) are generated by symmetry actions $s$ and $r$ on a triangle, where s is a rotation by $\pi$ radians about an axis connecting the center and a vertex of the triangle, and r the rotation around the center of the triangle by $2\pi/3$. Conjugacy classes of the group are thus given by $[e]=\{e\}$, $[s]=\{s^1r^1,s^1r^0,s^1r^2\}$ and $[r]=\{s^0r^1,s^0r^2\}$. The centralizers of the conjugacy classes are given by $N_{[e]}=S_3$ with three representations $0,1,2$, $N_{[r]}\cong \Z_3$ with three representations $0,1,2$, $N_{[r]}\cong \Z_2$ with three representations $0,1$. There are eight basic ground states introduced in the table below, following the notation of ref.\cite{Full}.

\begin{center}
\begin{table}[htbp]
\label{s3}
\begin{tabular}{ |c|c|c|c|c| } 
\hline $\mathcal{J}$ & 1 & 2 & 3 & 4   \\ 
\hline  $(A,h^A)$ & $([e],0)$ & $([e],1)$ & $([e],2)$ & $([r],0)$\\ 
\hline $d$ & 1 & 1 & 2 & 2\\
\hline  $\theta$ & 1 & 1 & 1 & 1  \\ 
\hline  
\hline
\hline $\mathcal{J}$ & 5 & 6 & 7 & 8 \\
\hline $(A,h^A)$ &  $([r],1)$ & $([r],2)$ & $([s],0)$ & $([s],1)$\\ 
\hline $d$ & 2 & 2 & 3 & 3\\
\hline $\theta$ & $e^{2\pi i/3}$ & $e^{-2\pi i/3}$ & 1 & -1 \\
\hline
\end{tabular}
\caption{Notations, quantum dimensions and twists for elementary ground states of model $Rep_{S_3}$.}
\end{table}
\end{center}

We have the nonzero entries of the decomposition matrix as,\begin{equation}
\begin{split}
& M_{10}=1,\ M_{21}=1,\ M_{32}=1, \\
& M_{40}=M_{41}=1,\ M_{52}=M_{62}=1,\\
& M_{70}=M_{72}=1,\ M_{80}=M_{82}=1.
\end{split}
\end{equation}

For a simplest graph on torus, the ground states can be written by,
\begin{equation}
\begin{split}
&\ket{1}=\frac{1}{\sqrt{6}}\left(\ket{000}+\ket{011}+2\ket{022}   \right);\\
&\ket{2}=\frac{1}{\sqrt{6}}\left(\ket{101}+\ket{110}-2\ket{122}   \right);\\
&\ket{3}=\frac{1}{\sqrt{6}}\left(\ket{202}-\ket{212}+\ket{220}-\ket{221}+\sqrt{2}\ket{222}   \right);\\
&\ket{4}=\frac{1}{\sqrt{6}}\left( \ket{000}+\ket{011}-\ket{022}+\ket{101}+\ket{110}+\ket{122}  \right);\\
&\ket{5}=\frac{1}{\sqrt{6}}(\ket{202}-\ket{212}+e^{-2\pi i/3}\ket{220}+e^{i\pi/3}\ket{221}\\
& \ \ \ \  +\sqrt{2}e^{2\pi i/3}\ket{222});\\
&\ket{6}=\frac{1}{\sqrt{6}}(\ket{202}-\ket{212}+e^{2\pi i/3}\ket{220}+e^{-\pi i/3}\ket{221}\\
& \ \ \ \  +\sqrt{2}e^{-2\pi i/3}\ket{222});\\
&\ket{7}=\frac{1}{\sqrt{6}}\left(\ket{000}-\ket{011}+\ket{202}+\ket{212}+\ket{220}+\ket{2211}   \right);\\
&\ket{8}=\frac{1}{\sqrt{6}}\left(\ket{101}+\ket{110}+\ket{202}+\ket{212}-\ket{220}-\ket{221}   \right).\\
\end{split}
\end{equation}

Again direct computations to $L_1+L_2=9$ show the above formula \eqref{SL1L2} for entanglement entropy to be exact.

\section{Summary and Discussions}\label{Summary}
In this work we studied quantum entanglement in topological phases on torus. We proposed a general formula for reduced density matrix \eqref{rdmMix} and entanglement entropy in string-net models on torus \eqref{SL1L2} with cylindrical partitions. We studied the minimal entanglement and generalized minimally entangled states to minimally entangled sectors. 

With the diagonalized reduced density matrices explicitly obtained, one can easily calculate the entanglement spectra and R\'enyi entropies of relevant systems.
General R\'enyi entropy is defined as:
\begin{equation}
S_R=\frac{1}{1-\alpha'}\log\ tr \left(\overline{\rho}_A^{\alpha'}\right).
\end{equation}

For all the examples mentioned above, we found the topological entanglement entropy unchanged with respect to $\alpha'$ in the large-$L$ limit, while the $L$-dependent term varies with $\alpha'$. This independence of $\alpha'$ in topological R\'enyi entropy is due to the simple structure of our reduced density matrix. Furthermore, increase in $\alpha$ leads to severer finite-size effects and takes larger $L_1, L_2$ for the entanglement entropy to grow stably in a linear way, which is consistent with previous results\cite{Renyi}.

One direct future direction will be to look at the general relationship between boundary theories and entanglement entropy because of the similar mathematical structure of our decomposition matrix and that of anyon condensations\cite{Bais1,Bais3,Hung,Kapustin,Levin,Barkeshli,KK,Kong}. 

It turns out that partial trace in the calculation of entanglement entropies corresponds to the condensation of all fluxes. To see this, we use the $Rep_{\Z_2}$ toric code\cite{Toric} model as an example. We have topological charges of the ground states $\J\in\left\{\unit,e,m,\epsilon\right\}$ with $\unit$ and $m$ the fluxes, while $j\in \left\{0,1\right\}$. When these topological charges move to the boundary, the fluxes are traced out, which are described by the action of
$(\unit \oplus m)$:

\begin{equation}
\label{eq:Z2}
\begin{split}
& \unit \otimes (\unit \oplus m) = (\unit \oplus m),\\
& m \otimes (\unit \oplus m) = (\unit \oplus m),\\
& e \otimes (\unit \oplus m) = (e \oplus \epsilon),\\
& \epsilon \otimes (\unit \oplus m) = (e \oplus \epsilon).\\
\end{split}
\end{equation}

We observe that upon fusion with fluxes, the four topological charges of the ground states fall into two entanglement sectors. On the right hand side of above equations, we relabel $(\unit \oplus m)$ as $a$ and $(e \oplus \epsilon)$ with $b$ and view them as simple objects $a,b\in I$. One can figure out the fusion rules between $a$ and $b$ making use of a \textit{principle: moving and fusion commutes}. Namely, if we pick $m$ and $e$, fuse them as in the bulk, we obtain $\epsilon$. Now we move this fusion result to the boundary of the cuts and do the partial trace, it becomes $b$. If instead, we pick $m$ and $e$, first move them separately to the boundary and do the partial trace, they become $a$ and $b$ separately. Then we can fuse $a$ and $b$. The principle says the final results of these two different processes must be the same, i.e., $a\otimes b=b$. We can pick other $\J$'s and obtain all the fusion rules between $a$ and $b$: \begin{equation}
a\otimes a=a,\ a\otimes b=b\otimes a=b,\ b\otimes b=a.\nonumber
\end{equation}
These exactly corresponds to the fusion rules of the $Rep_{\Z_2}$ model, thus we identify $a=0$, $b=1$. With this identification, we directly arrive at the decomposition matrix \eqref{M:Z2}. Similar arguments apply for calculation of $M$ matrices for more complicated input data. Furthermore, the topological charges of ground states that are in the same entanglement sector, when interpreted as bulk quasiparticles, map to the same excitations on the boundary. 
 
These similarities correspond to the mathematical fact that the functor for partial trace $\mathcal{F}:\mathcal{Z}(\mathcal{C})\rightarrow \mathcal{C}$ is a special case of the bulk-to-boundary map $\mathcal{F}:\mathcal{Z}(\mathcal{C})\rightarrow \mathcal{D}$ with $\mathcal{D}$ satisfying $\mathcal{Z}(\mathcal{C})\simeq \mathcal{Z}(\mathcal{D})$. It would be interesting to see the physical meaning of this correspondence.

We note for clarity that our decomposition $M_{\J j}$ matrix is different from the tunneling matrices in the context of gapped domain walls studied in ref.\cite{Lan,Bernevig2}. There $j$ is a boundary excitation in the only the deconfined part of the anyon-condensed state. In contrast, the $j$ in our expressions labels the boundary degrees of freedom along our partitioning cuts, or a boundary excitation in the whole anyon-condensed state, including the confined ones. Furthermore, the boundaries in our context are not real physical boundaries, but boundaries of the partitions. 

Our results on nonchiral lattice systems are different from the those in chiral cases studied in ref.\cite{Fradkin,Ashvin}. We expect this difference to be essential and believe that future studies on the relationship between boundary theory and entanglement will provide deeper insights to the issue.

Finally, it is a curious question whether the algebra structure of $Hom(q,q)$ introduced in Appendix \ref{app:Lemma} will hold for more general lattice spin systems. If this is the case, then the above discussions can possibly imply general mathematical structure of entanglement.

\begin{acknowledgments}
We thank Yang Qi for valuable comments and suggestions. Zhuxi thanks Ren Pankovich for inspiring discussions.
\end{acknowledgments}

\appendix
\section{String-net Model}
\label{SN}
We briefly review the string-net model\cite{StringNet} in this appendix. The input data $\left\{I, d, \delta, G\right\}$ to define the model comes from a unitary fusion category $\mathcal{C}$.

The model is defined on a trivalent graph on a closed oriented surface. Degrees of freedom live on links of the graph. For each link, we assign a string type $j\in I=\{j=0,1,...,N\}$, where $I$ is called the label set. In the case of lattice gauge theories, $j$'s label the irreducible representations of a group. More generally, they can label irreducible representations of quantum groups. The Hilbert space is spanned by all configurations of the labels on links. Each label $j$ has a ``conjugate'' $j^*\in I$, satisfying $j^{**}=j$. There is unique  ``vacuum'' label $j=0$ with $0^*=0$. We require the state to be the same if one reverses the direction of one link and replaces the label $j$ by $j^*$, which is a graphical realization of time reversal symmetry. 

We associate to each string type a number $d_j$ called quantum dimension of $j$, and define the total quantum dimension to be $D=\sum_j d_j^2$. We further assign to each three string types a tensor $\delta_{ijk}$ which specifies the branching rules of a trivalent graph. If for some $i,j,k\in I$ one has $\delta_{ijk}=1$, then the three string types are allowed to meet at a vertex. Otherwise their meeting is not energetically favored, i.e., we will have charge excitations on the corresponding vertex.

Given the quantum dimensions and fusion rules, we define the symmetrized $6j$-symbols, often denoted as $G$. They are complex numbers satisfying the following conditions\cite{Full}:
\begin{align}
\label{eq:6jcond}
\begin{array}{ll}
&G^{ijm}_{kln}=G^{mij}_{nk^{*}l^{*}}
=G^{klm^{*}}_{ijn^{*}}=\alpha_m\alpha_n\,\overline{G^{j^*i^*m^*}_{l^*k^*n}},\\
&\sum_{n}{d_{n}}G^{mlq}_{kp^{*}n}G^{jip}_{mns^{*}}G^{js^{*}n}_{lkr^{*}}
=G^{jip}_{q^{*}kr^{*}}G^{riq^{*}}_{mls^{*}},\\
&\sum_{n}{d_{n}}G^{mlq}_{kp^{*}n}G^{l^{*}m^{*}i^{*}}_{pk^{*}n}
=\frac{\delta_{iq}}{d_{i}}\delta_{mlq}\delta_{k^{*}ip},\\
\end{array}
\end{align}
where the first condition specifies tetrahedral symmetry, the second the pentagon identity, and the third orthogonality condition. The number $\alpha_j$ is the Frobenius-Schur indicator. In the example of lattice gauge theories, this indicator tells whether the representation $j$ is real, complex, or pseudoreal. In this case, $d_j=\alpha_j\mathrm{dim}(j)$ with $\mathrm{dim}(j)$ the corresponding dimension of the space of the representation $j$; and the tensor $G_{kln}^{ijm}$ the (symmetrized) Racah $6j$ symbol for the group. In this example, string-net model can be mapped to the Kitaev's quantum double model. 

There are two types of local operators that are used to specify the Hamiltonian. On every vertex $v$, we have ${Q}_v=\delta_{ijk}$ acting on the labels of three edges incoming to the vertex $v$. On every plaquette $p$, we have $B_p^s$ with $s\in I$. It acts on the boundary edges of the plaquette $p$, and has the matrix elements on a triangle plaquette\cite{Full},
\begin{align}
\label{eq:Bps::InLW}
&\Biggl\langle
\TriangleYpart{j_4}{j^{\prime}_1}{j^{\prime}_2}{j^{\prime}_3}{j_5}{j_6}{>}{<}{>}{<}{<}{<}
\Biggr|
B_p^s
\Biggl|\TriangleYpart{j_4}{j_1}{j_2}{j_3}{j_5}{j_6}{>}{<}{>}{<}{<}{<}\Biggr\rangle\nonumber\\
=&
v_{j_1}v_{j_2}v_{j_3}v_{j'_1}v_{j'_2}v_{j'_3}
G^{j_5j^*_1j_3}_{sj'_3j^{\prime*}_1}G^{j_4j^*_2j_1}_{sj'_1j^{\prime*}_2}G^{j_6j^*_3j_2}_{sj'_2j^{\prime*}_3},
\end{align}
where $v_j=\sqrt{d_j}=\frac{1}{G^{j^*j0}_{0\,0\,j}}$. The same rule for $B_p^s$ applies when the plaquette $p$ is a quadrangle, a pentagon, etc.. 

Defining $B_p=\frac{1}{D}\sum_{s}d_sB_p^{s}$, we have ${Q}_v$ and $B_p $'s mutually-commuting: $[Q_v,Q_{v^{\prime}}]=0=[B_p,B_{p^{\prime}}],[Q_v,B_p]=0$; furthermore, they are also projectors: ${Q}_v^2={Q}_v$ and $B_p^2=B_p$.

The Hamiltonian of the model is
\begin{equation}
  \label{eq:HamiltonianLW}
  {H}=-\sum_{v}{Q}_v-\sum_{p}B_p,  
\end{equation}
where the sum run over vertices $v$ and plaquettes $p$ of the whole trivalent graph. Because of the commutative property of $Q_v$ and $B_p$'s, the Hamiltonian is exactly soluble. Ground states satisfies ${Q}_v=B_p=1$ for all $v$,$p$.

\section{Fiber Fusion Category}\label{app:Fiber}
A fusion category can be uniquely defined from a fusion system\cite{FCC} $\left\{I,d,\delta,G\right\}$. Given the input data of string-net model, one can construct a fiber tensor category $\mathcal{F}(I,d,\delta,G)$, where
\begin{itemize}
\item[(1)] Objects are vector bundles over $I$, specified by $n: I\rightarrow \mathbb{N}$, with $\mathbb{N}$ the set of non-negative integers. An object in the category is $\left\{\mathbb{C}^{n_j}\right\}_{j\in I}$.
\item[(2)] Morphisms $f: m\rightarrow n$ are sets of linear maps $\left\{f_j: \mathbb{C}^{m_j}\rightarrow \mathbb{C}^{n_j}\right\}_{j\in I}$. We denote the hom space (the set of all morphisms mapping $m$ to $n$) $Hom(m,n)=\oplus_{j\in I} Mat_{n_jm_j}$ and abbreviate it by $Mat_{nm}$. 
\item[(3)] Tensor product between objects $m\otimes n$ is defined by, \begin{equation}
\left(m\otimes n\right)_k = \sum_{i,j\in I} m_i n_j \delta_{ijk^*}.
\end{equation}
\item[(4)] Tensor product between morphisms: 
\begin{equation}
\left(f\otimes g\right)_k=\mathop{\oplus}\limits_{i,j\in I}\left(f_i\otimes g_j\right)\delta_{ijk^*}\in Mat_{(m\otimes n)_k,(p\otimes q)_k}.
\end{equation}
\item[(5)] There are further consistency conditions needed to be satisfied which are given by the normalized $6j$-symbols G as in the Appendix \ref{SN}.
\end{itemize}

Consider an example from the Fibonacci category. The label set is $I=\left\{0,1\right\}$ (or $\left\{\unit,\tau\right\}$ in usual notation), all the strings are self-dual and the nontrivial fusion rule is $\delta_{110}=\delta_{111}=1$.

Let us consider an object corresponding to $Q=\oplus_{j\in I}\ j$, denoted by $q: I\rightarrow \left\{1\right\}$. Intuitively, this is assigning to both $j=0,1$ a ``fiber'' $\mathbb{C}$. The simple object corresponding to $q$ is $\left\{\mathbb{C}\right\}_{j\in I}$. The tensor product of morphisms is 
\begin{equation}
\begin{split}
& \left(q\otimes q\right)_0=\delta_{000}+\delta_{110}=2,\\
& \left(q\otimes q\right)_1=\delta_{011}+\delta_{101}+\delta_{111}=3,\nonumber
\end{split}
\end{equation}
as shown in the figure below.

\begin{figure}[htbp]
\centering
\begin{tikzpicture}
\draw (0,0)--(3,0);
\draw (1,0)--(2,1);
\draw (2,0)--(3,1);
\node[below] at (1,0) {$0$};
\node[below] at (2,0) {$1$};
\node[right] at (2,0.8) {$\mathbb{C}$};
\node[right] at (3,0.8) {$\mathbb{C}$};
\draw (5,0)--(8,0);
\draw (6,0)--(7,1);
\draw (7,0)--(8,1);
\node[below] at (6,0) {$0$};
\node[below] at (7,0) {$1$};
\node[right] at (7,0.8) {$\mathbb{C}^2$};
\node[right] at (8,0.8) {$\mathbb{C}^3$};
\end{tikzpicture}
\caption{Left: example of a simple object in fiber fusion category. Right: example of the tensor product of the two same simple objects on the left.}
\end{figure}
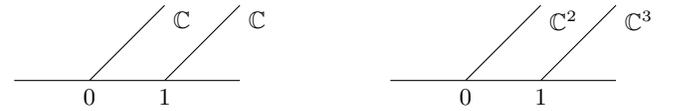

A morphism $\alpha\in Hom(q,q)=\oplus Mat_{11}$ is just a complex function over $I$. In the matrix form, this is \begin{equation}\alpha=\left(
\begin{matrix}
& \alpha_0 &   \\
&   & \alpha_1 \\
\end{matrix}\right).\nonumber
\end{equation}
Then tensor product of two morphisms $\alpha,\beta \in Hom(q,q)$ has the matrix form \begin{equation}\alpha\otimes\beta=\left(
\begin{matrix}
& \left(\begin{matrix}
& \alpha_0\beta_0 &    \\
&   & \alpha_1 \alpha_1 \\
\end{matrix}\right) &  \\
&   & \left( \begin{matrix}
& \alpha_0\beta_1 &   &  \\
&   & \alpha_1\beta_0 & \\
&   &   & \alpha_1\beta_1\\
\end{matrix}\right)
\end{matrix}\right).\nonumber
\end{equation}

\section{Proofs}\label{app:Lemma}
Proofs are based on fiber fusion category Appendix \ref{app:Fiber}.

\subsection{Proof of Lemma (1)} 
To prove eq.\eqref{eq:Lemma1}, consider an algebra of $Hom(q,q)$ with multiplication ``$\cdot$'' given by
\begin{equation}
(f\cdot g)_j=tr\left((f\otimes g)_j\right).\nonumber
\end{equation}
Then $\alpha\cdot\alpha=\alpha$. Furthermore, the trace becomes  \begin{equation}
tr \beta_j = tr \left( P_j \left(\alpha^{\cdot L}\right)\right)=\alpha_j.\square \nonumber
\end{equation}

\subsection{Proof of Lemma (2)} 
The identity to be proved is eq.\eqref{eq:Lemma2}. Using Lemma (1) and also the identity $\sum_i \frac{d_i^2}{D}=1$, the proof simplifies to the verification of 
\begin{equation}
\label{eq2}
tr\left(\mathop{\oplus}\limits_{i} \frac{d_i}{D} \beta_i \log \beta_i\right)-tr\left(\beta_0 \log \beta_0\right)=0.\nonumber
\end{equation}

Since $L\geq 2$, we can rewrite $\beta=\sigma\otimes \alpha$, where $\sigma=\alpha^{\otimes (L-1)}$. Then the above equation becomes,

\begin{widetext}
\begin{equation}
\begin{split}
& tr\left[\mathop{\oplus}\limits_{i}\frac{d_i}{D} \left(\sigma\otimes \alpha\right)_i \log \left(\sigma\otimes \alpha\right)_i\right]-tr\left[\left(\sigma\otimes \alpha\right)_0 \log \left(\sigma\otimes \alpha\right)_0\right] \\
=& \sum_i \frac{d_i}{D} \sum_{j,k} tr\left[\left( \sigma_j \frac{d_k}{D}\delta_{jki^*}\right)\log \left( \sigma_j \frac{d_k}{D}\right)\right]-\sum_{j,k} tr\left[\left( \sigma_j \frac{d_k}{D}\delta_{jk0^*}\right)\log \left( \sigma_j \frac{d_k}{D}\right)\right]\\
=& \left(\sum_{ijk} \frac{d_id_k}{D^2}\delta_{jki^*}tr\left(\sigma_j\log \sigma_j\right)-\sum_j\frac{d_j}{D}tr\left(\sigma_j\log \sigma_j\right)\right)+\left(\sum_{ijk} \frac{d_id_j}{D^2}\delta_{jki^*}\frac{d_k}{D}\log \frac{d_k}{D}-\sum_j \frac{d_j^2}{D^2}\log \frac{d_j}{D}\right)\\
=& 0 +\sum_k\frac{d_k}{D^2}\log \frac{d_k}{D}\sum_{ij} \frac{d_id_j}{D}\delta_{ijk^*}-\sum_j \frac{d_j^2}{D^2}\log \frac{d_j}{D}\\
=& 0.\nonumber
\end{split}
\end{equation}
\end{widetext}
Here in the second equality we have used the condition that the pairs $i,k$ satisfying $\delta_{jk0^*}=1$ also satisfies $\delta_{0k^*j^*}=\delta_{k^*0j^*}=\delta_{j0k}=1$. Thus $j=k^*$. Lemma (1) and the identity $Dd_j=\sum_{ik}d_id_k\delta_{ikj^*}$ are also used in the derivation. $\square$.

\section{Half-braiding $z$-Tensor}
\label{app:HalfBraiding}

Below we give a brief review of half-braiding tensors summarized from ref.\cite{Full,Thesis}.

Given the input data $\left\{I,d,\delta, G\right\}$, one can define the half-braiding $z$ tensor by the following naturality condition for all $ p, q, j, k, t, m, n.\in I$:

\begin{equation}
\sum_{lrs}d_rd_sz_{lnqr}z_{pmls}G^{m^*sl^*}_{nr^*t}G^{s^*pm}_{jn^*t}G^{m^*tr^*}_{q^*n^*k}=\delta_{j,k}\delta_{mnj^*}\frac{1}{d_j}z_{pjqt}.
\end{equation}

We enumerate every nonzero irreducible solution of $z_{pjqt}^{\J}$ by a label $\J$, being irreducible in the sense that it cannot be further decomposed into linear combinations of some $z_{pjqt}^{\J_1}, z_{pjqt}^{\J_2}, \dots$ that also satisfy the naturality conditions. The algebraic theory of all these irreducible solutions is the quantum (Drinfeld) double theory of the input data, where each $\J$ corresponds to an irreducible representation of the tube algebra\cite{Muger, Full} (also appearing in a form called   $Q$-algebra\cite{Qalgebra}). 

Quantum double labels classify species of bulk quasiparticles. For each $\J$, we define the twist from any $j\in I$ that satisfies $M_{\J j}=1$:
\begin{equation}
\theta_\J=\frac{1}{d_j M_{\J j}}\sum_t z_{jjjt}^\J d_t \delta_{jjt^*}.
\end{equation}

The $S$ matrix is,
\begin{equation}
S_{\J\mathcal{K}}=\frac{1}{D}\sum_{i,j,k\in I}d_kz_{ijik}^\J z_{jijk}^\mathcal{K}.
\end{equation}

We remark that $z^J_{pjqt}$ appears in the original string-net paper\cite{StringNet} as the $\Omega$ tensor, with a slightly different normalization.

\newpage
\nocite{*}
\bibliographystyle{unsrt}
\bibliography{bibliography}

\end{document}